%% file: root.tex
\documentclass[letterpaper, 10 pt, conference]{ieeeconf}  

\IEEEoverridecommandlockouts                              

\usepackage{textcomp}
\usepackage{stfloats}
\usepackage{url}
\usepackage{verbatim}
\usepackage{graphicx}
\usepackage[sort,compress,nospace]{cite}
\usepackage{threeparttable}
\usepackage{tablefootnote}
\usepackage[dvipsnames]{xcolor}
\usepackage{amsmath,amsfonts}
\usepackage{algorithmic}
\usepackage{algorithm}
\usepackage{array}
\usepackage{siunitx}
\usepackage{epstopdf}
\usepackage[caption=false, font=footnotesize]{subfig}
\usepackage{svg}
\usepackage{multirow}
\usepackage{booktabs}

\title{\LARGE \bf
Inferring Occluded Agent Behavior in Dynamic Games\\ from Noise Corrupted Observations
}

\author{Tianyu Qiu$^{1}$ and David Fridovich-Keil$^{1}$
\thanks{$^{1}$Tianyu Qiu and D. Fridovich-Keil are with the Aerospace Engineering department, University of Texas at Austin. This research is supported by the National Science Foundation under grant 2211548. {\tt\small tianyuqiu@utexas.edu, dfk@utexas.edu}}
}

\begin{document}

\maketitle
\thispagestyle{empty}
\pagestyle{empty}

\begin{abstract}

In mobile robotics and autonomous driving, it is natural to model agent interactions as the Nash equilibrium of a noncooperative, dynamic game. These methods inherently rely on observations from sensors such as lidars and cameras to identify agents participating in the game and, therefore, have difficulty when some agents are occluded. To address this limitation, this paper presents an occlusion-aware game-theoretic inference method to estimate the locations of potentially occluded agents, and simultaneously infer the intentions of both visible and occluded agents, which best accounts for the observations of visible agents. Additionally, we propose a receding horizon planning strategy based on an occlusion-aware contingency game designed to navigate in scenarios with potentially occluded agents. Monte Carlo simulations validate our approach, demonstrating that it accurately estimates the game model and trajectories for both visible and occluded agents using noisy observations of visible agents. Our planning pipeline significantly enhances navigation safety when compared to occlusion-ignorant baseline as well.
\end{abstract}

\section{Introduction}
\input{introduction}

\section{Related Work}
\input{related_work}

\section{Preliminaries}\label{Preliminaries}
\input{preliminaries}

\section{Methods}
\input{method}

\section{Simulation Experiments}
\input{experiment}

\section{Conclusion \& Future Work}
\input{conclusion}

\bibliographystyle{IEEEtran}
\bibliography{IEEEabrv, reference.bib}
\end{document}

%% file: introduction.tex
\begin{figure}[!t]
\centering
\includegraphics[width=0.47\textwidth]{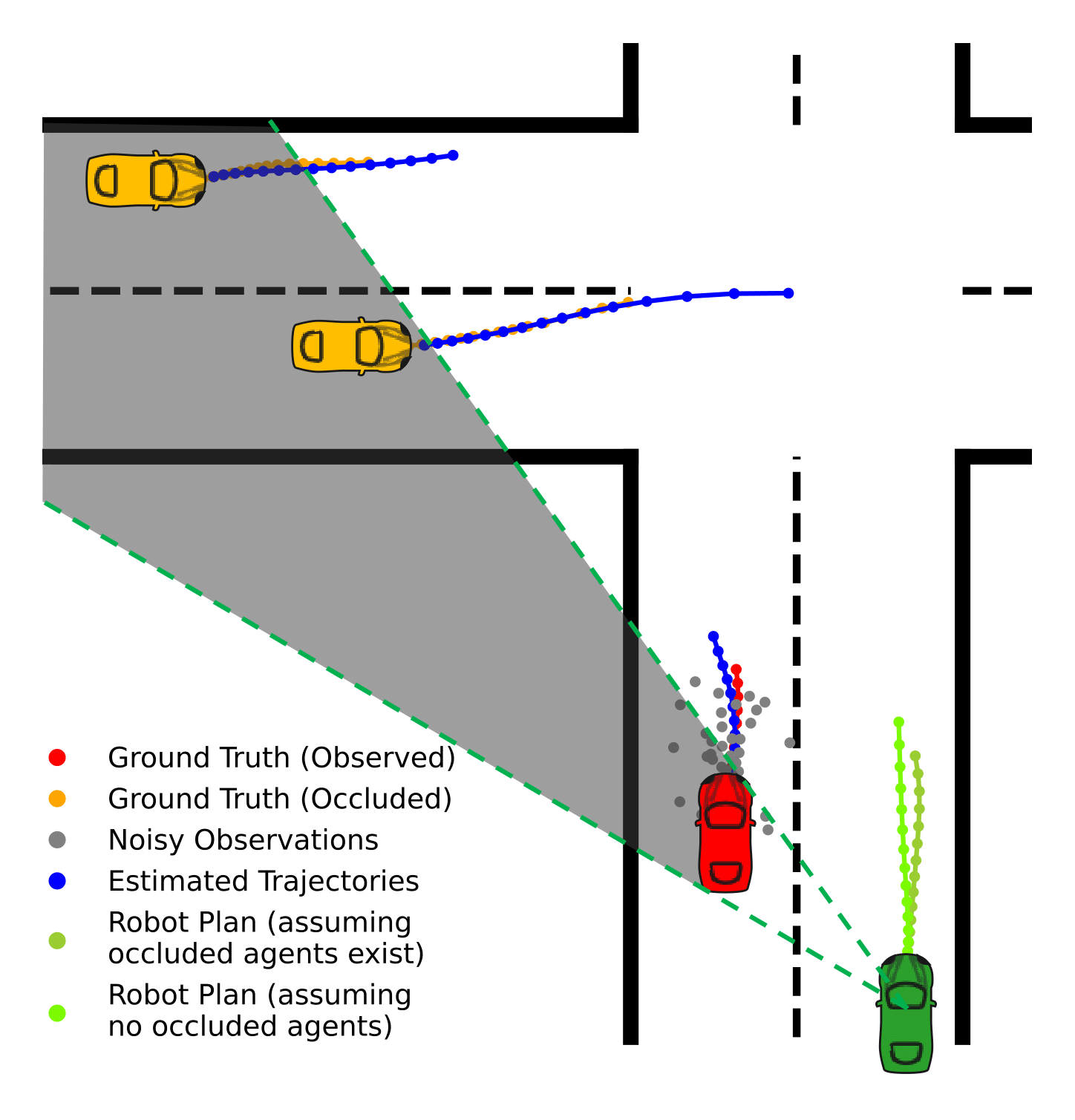}
\caption{Occlusion-aware contingency game planner in an intersection scenario. The green vehicle can only see the red vehicle in the adjacent lane and is uncertain about the existence of occluded vehicles in the horizontal lanes.
The green vehicles makes two assumptions: either 1) occluded vehicles exist. It can then use our proposed approach to estimate their trajectories by observing the red vehicle’s deceleration and plans the dark green trajectory; or 2) there are no occluded vehicles. It then only interacts with the red vehicle and plans the light green trajectory. Our contingency planning approach blends these two alternative strategies, while accounting for the fact that any occluded agents will be visible in the near future as the green vehicle approaches the intersection.}
\label{fig:front_figure}
\vspace{-4ex}
\end{figure}

Robots extensively rely on sensor observations to detect static and dynamic obstacles. However, sensors inherently face limitations, primarily due to occlusion or range constraints. For instance, consider a common scenario illustrated in Figure \ref{fig:front_figure} where a green vehicle can only observe a 
nearby red vehicle, which occludes the orange vehicles in the horizontal lanes. If the green vehicle maintains its speed, it risks colliding with the occluded vehicles. To navigate safely in such conditions, the green vehicle can consider two alternatives: 1) no occluded vehicles exist, making it safe to drive forward, and 2) occluded vehicles exist, which is especially likely if the green vehicle observes the red vehicle decelerating. 
A responsible human driver would choose to proceed cautiously, balancing the risk of collision with an occluded vehicle against the (perhaps more likely) scenario in which it is safe to proceed. A reasonable driver would also recognize that the horizontal lanes will not remain occluded for very long, at which point any occluded vehicles will be visible and the optimal behavior will be unambiguous.

This scenario highlights how interactions among agents influence their observed behavior and, conversely, how observed behavior can reveal underlying interaction structures. For example, the deceleration of the red vehicle suggests the existence of occluded orange vehicles. Therefore, robots can anticipate and respond to potential occluded agents promptly by observing the behavior of visible agents.

Existing game-theoretic methods
effectively solve multi-agent interaction and planning problems 
by computing agents' trajectories as Nash
equilibria of noncooperative, dynamic games \cite{fisac2019hierarchical,fridovich2020efficient,laine2021multi,chiu2021encoding,wang2021game,le2022algames,laine2023computation,zhu2024sequential}.
Recent efforts are also capable of identifying the game model that best describes the observed agent behaviors \cite{le2021lucidgames,peters2021inferring,peters2023ijrr,li2023cost, liu2023learning,mehr2023maximum, khan2024leadership}. These methods perform well when all agents are perfectly visible. However, the presence of occluded agents, a common occurrence in real-world scenarios, has been largely overlooked.

To address the challenge of occluded agents, we make the following contributions which are illustrated in Figure~\ref{fig:front_figure}: 1) an occlusion-aware game-theoretic estimator that simultaneously identifies unknown parameters in each agent’s cost function and estimates the trajectories of both occluded and visible agents in a Nash game, based solely on noisy position observations of visible agents; and 2) a receding horizon occlusion-aware contingency game planner, structured to account for the possible existence or nonexistence of occluded agents. We evaluate our method across various traffic scenarios.
Monte Carlo experiments indicate that our estimation approach provides accurate trajectory estimates for both visible and occluded agents. Moreover, our planning pipeline computes safer navigation decisions compared to the existing baseline method.

%% file: related_work.tex
\subsection{Social Occlusion Inference}\label{soi_intro}
We begin by introducing social occlusion inference methods. Many works \cite{hara2018recognizing,hara2020predicting, afolabi2018people,itkina2022multi,mun2023occlusion} have explored inference methods based on observations of visible agents' behavior in social settings. These works leverage the influence of the surrounding environment on an individual's behavior. Representative works often employ computer vision techniques \cite{hara2018recognizing,hara2020predicting} and occupancy grid map generation techniques \cite{afolabi2018people,itkina2022multi,mun2023occlusion} to infer occupancy of occluded areas. While these methods primarily focus on determining whether the occluded area is occupied or if an agent is emerging from it, additional techniques must be applied to provide precise information about agents' behavior or to compute safe control policies for robots\cite{mun2023occlusion}.

\subsection{Planning-based Agent Behavior Modeling}\label{planning_intro}
To enable precise behavior modeling, many works utilize planning-based methods. These approaches assume that agents make rational decisions about their trajectories, which follow specific rules. Inferring these rules is the key to behavior modeling tasks. Inverse optimal control (IOC) \cite{kalman1963linear,maroger2020walking,albrecht2011imitating, maroger2022inverse} and inverse reinforcement learning (IRL)\cite{ng2000algorithms,ziebart2009planning,kitani2012activity,monfort2015intent,previtali2016predicting} are two prominent frameworks in this area. These approaches focus on recovering the rules that underlie agent-environment interactions and use these rules to predict behavior. Recent multi-agent IRL techniques \cite{henry2010learning,pfeiffer2016predicting,kim2016socially,kretzschmar2016socially,morales2018towards,sun2019behavior} explore rules governing agents' social behaviors and apply to robot navigation \cite{kim2016socially,kretzschmar2016socially} and autonomous driving \cite{morales2018towards,sun2019behavior} tasks.

\subsection{Dynamic Games and Inverse Dynamic Games}\label{game_intro}
As a subcategory of planning-based methods, game-theoretic methods\cite{fisac2019hierarchical,fridovich2020efficient,laine2021multi,chiu2021encoding,wang2021game,le2022algames,laine2023computation,zhu2024sequential} extend single-agent optimal control techniques to model multi-agent interactions as Nash equilibria of noncooperative, dynamic games\cite{bacsar1998dynamic}. Recent efforts \cite{le2021lucidgames,peters2021inferring,peters2023ijrr,li2023cost,liu2023learning,mehr2023maximum} aim to model agent behaviors more precisely by introducing inverse dynamic games. Instead of directly computing Nash trajectories, inverse games aim to learn the parameters of a game model that best explains the observed agents' behavior.

In general, planning-based and game-theoretic methods assume that agents' behaviors directly result from 
their attempts to optimize specific preferences which are encoded as objective functions.
However, these approaches share a common limitation: they require observations from all agents and do not account for occluded agents, leading to imprecise inference of visible agents' objectives and suboptimal navigation outcomes in the presence of occlusions. 
To address this challenge, we propose an occlusion-aware game-theoretic technique for estimating both observed and occluded agents' state and intentions, from only observations of visible agents. 

\subsection{Contingency Planning and Contingency Games}
Contingency planning has been studied extensively as a technique for managing uncertainty in navigation tasks. Contingency planning methods \cite{hardy2013contingency, chen2022interactive, nair2022stochastic} design a control strategy with multiple branches representing both the different possible scenarios and the times at which an agent would be sure of which scenario it is in.
Recent works \cite{rhinehart2021contingencies, tolstaya2021identifying} also propose an end-to-end contingency planning framework that enables both intention prediction and motion planning. In multi-agent scenarios, contingency games can additionally account for agents' noncooperative behavior \cite{peters2024contingency}.

In dealing with potentially occluded agents during navigation, our method extends the idea in \cite{peters2024contingency} by proposing an occlusion-aware contingency game planner. We introduce a receding-horizon estimation and planning pipeline that enhances accuracy in modeling agent behavior and promotes safety in scenarios with potentially occluded agents.

%% file: preliminaries.tex
\begin{table}[H]
\caption{Main Symbols and Notations\label{tab:table1}}
\centering
\begin{tabular}{l|l}
\hline
${J^i}$ & Cost function (objective) of the ${i^{\text{th}}}$ agent\\
${M\in \mathbb{N}^*}$ & Number of agents in the game\\
${\textbf{w}^i=(w_1^i,w_2^i,\dots)}$ & Weighting parameters of the ${i^{\text{th}}}$ agent\\
${\textbf{w}=(\textbf{w}^1,\dots,\textbf{w}^M)}$ & Tuple of all agents' weighting parameters\\
${T\in \mathbb{N}^*}$ & Planning horizon of the game\\
${K\in \mathbb{N}^*}$ & Observation interval of the game\\
$\Delta$ & Time interval between two adjacent steps\\
${x_k^i\in\mathbb{R}^n}$ & ${i^{\text{th}}}$ agent's state at time ${k}$\\
${\textbf{x}^i:=(x_1^i,\dots,x_T^i)}$ & ${i^{\text{th}}}$ agent's trajectory\\
${\textbf{x}_k:=(x_k^1,\dots,x_k^M)}$ & All agents' state at time ${k}$\\
${\textbf{x}:=(\textbf{x}^1,\dots,\textbf{x}^M)}$ & Tuple of all agents' trajectories\\
${u_k^i\in\mathbb{R}^m}$ & ${i^{\text{th}}}$ agent's control input at time ${k}$\\
${\textbf{u}^i:=(u_1^i,\dots,u_T^i)}$ & ${i^{\text{th}}}$ agent's control sequence\\
${\textbf{u}_k:=(u_k^1,\dots,u_k^M)}$ & All agents' control input at time ${k}$\\
${\textbf{u}:=(\textbf{u}^1,\dots,\textbf{u}^M)}$ & Tuple of all agents' control sequences\\
${\mathcal{L}^i}$ & Lagrangian of the ${i^{\text{th}}}$ agent\\
${\lambda_k^i\in\mathbb{R}^n}$ & ${i^{\text{th}}}$ agent's Lagrange multiplier at time ${k}$\\
${\pmb{\lambda}^i:=(\lambda_1^i,\dots,\lambda_T^i)}$ & Sequence of ${\lambda_k^i}$ of the ${i^{\text{th}}}$ agent\\
${\mathcal{V}}\subseteq[M]$ & Set of visible agents in the game\\
${\mathcal{V}_i}\subseteq[M]$ & Set of agents visible to the $i^{\text{th}}$ agent\\
${\mathcal{O}}:=[M]\setminus\mathcal{V}$ & Set of occluded agents in the game\\
$\theta\in\Theta$ & Set of hypotheses $\theta$ in a contingency game\\
\hline
\end{tabular}
\end{table}

A discrete time open-loop Nash game with $M$ agents is characterized by state ${x_k^i\in\mathbb{R}^n}$ and control inputs ${u_k^i\in\mathbb{R}^m}$, ${i\in \{0,\cdots,M-1\} \equiv [M]}$,
at time step $k$. The dynamics ${x_{k+1}^i=f(x_k^i,u_k^i)}$ govern each agent's state transition from time $k$ to $k+1$. ${J^i(\textbf{x},\textbf{u};\textbf{w}^i):=\sum_{k=1}^K g_k^i(\textbf{x}_k,\textbf{u}_k;\textbf{w}^i)}$ defines the ${i^{\text{th}}}$ agent's weighted cumulative cost over a planning horizon $T$, and is parameterized by vector $\textbf{w}^i$. The game is thus fully characterized by the tuple of all agents' cost functions $\{J^i(\textbf{x},\textbf{u};\textbf{w}^i)\}_{i=1}^M$, the initial states ${\textbf{x}_0}$, and the dynamics $f$, and is denoted by ${\Gamma(\textbf{w},\textbf{x}_0,f)}$.
In this Nash game, each agent $i$ aims to minimize its cost function while satisfying dynamic feasibility constraints, i.e.
\begin{subequations}
    \begin{align}
    \min_{\textbf{x}^i,\textbf{u}^i}\ &J^i(\textbf{x},\textbf{u};\textbf{w}^i) 
    \label{eqn:nashobj}\\
    \text{s.t.}\ &x_{k+1}^i=f(x_k^i,u_k^i), \quad k\in[T].\label{eqn:nashdynamics}
    \end{align}
    \label{eqn:nashgame}%
\end{subequations}
\textbf{Open-Loop Nash Equilibrium}: If the inequalities
\begin{equation}
\label{nasheq}
    J^i(\textbf{x}^*,\textbf{u}^*;\textbf{w}^i)\leq J^i(\textbf{x}^i,\textbf{x}^{-i*},\textbf{u}^i,\textbf{u}^{-i*};\textbf{w}^i),\quad i\in[M],
\end{equation}
are satisfied for all $\textbf{x}^i, \textbf{u}^i$ that remain feasible with respect to \eqref{eqn:nashdynamics}, then $\textbf{u}^{i*}:=\{u_0^{i*},\cdots,u_{T-1}^{i*}\}$ is called a Nash strategy with ${\textbf{x}^{i*}:=\{x_1^{i*},\cdots,x_{T}^{i*}\}}$ being the corresponding open-loop Nash equilibrium (OLNE) trajectory. This inequality indicates that no agent can reduce their cost by unilaterally deviating from Nash strategy ${\textbf{u}^{i*}}$\cite{bacsar1998dynamic}.

We use the following example to concretize the concept.\\
\textbf{Running Example}: Consider a scenario with $M$ agents moving towards their goals while avoiding each other. At each time $k$, the $i^{\text{th}}$ agent's state $x_k^i$ encodes its position ${p_k^i=[p_{x,k}^i,p_{y,k}^i]^\top\in\mathbb{R}^2}$ and velocity ${v_k^i=[v_{x,k}^i,v_{y,k}^i]^\top\in\mathbb{R}^2}$ and evolves according to double-integrator dynamics, i.e.,
\begin{equation}
    \begin{aligned}
    x_{k+1}^i&=\begin{bmatrix}
            p_{k+1}^i\\
            v_{k+1}^i\\
        \end{bmatrix}=\begin{bmatrix}
            I_{2} & I_{2}\Delta\\
            \textbf{0} & I_{2}
        \end{bmatrix}\cdot \begin{bmatrix}
            p_{k}^i\\
            v_{k}^i\\
        \end{bmatrix}+\begin{bmatrix}
            \textbf{0} \\ 
            I_{2}
        \end{bmatrix}\Delta\cdot a_k^i\\
    &=Ax_k^i + Bu_k^i\Delta,\quad k\in[T],\ i\in[M].
    \end{aligned}
    \label{eqn:double-integrator dynamic}
\end{equation}%
where the control input ${u_k^i=a_k^i=[a_{x,k}^i,a_{y,k}^i]^\top}$ denotes its acceleration and $\Delta$ denotes the time interval between two adjacent steps. Its objective is to minimize the sum of the running cost ${g_k^i}$ over time, where ${g_k^i}$ is characterized by various features with non-negative weighting parameters ${\textbf{w}^i}$:
\begin{equation}
    g_k^i(\textbf{x}_k,\textbf{u}_k;\textbf{w}^i)=\sum_{l=1}^3w_l^ig_{l,k}^i\left\{\begin{aligned}
    &g_{1,k}^i=\|p_k^i-p_g^i\|_2^2\\
    &g_{2,k}^i=\sum_{\substack{j=1\\j\neq i}}^M\frac{1}{\|p_k^i-p_k^{j}\|_2^2}\\
    &g_{3,k}^i=\|u_k^i\|_2^2
    \end{aligned}\right.
    \label{eqn:running cost}
\end{equation}%
This form of objective guides the $i^{\text{th}}$ agent towards its destination $p_g^i$ (${g_{1,k}^i}$) while maintaining a safe distance from the other agents (${g_{2,k}^i}$) with minimal energy expenditure (${g_{3,k}^i}$). In practice, ${g_k^i}$ can be readily modified to accommodate different scenarios.

%% file: method.tex
\subsection{Occlusion-Aware Behavior Inference}
\label{sec:occluded_agent_behavior_inference}
We first introduce two roles in a dynamic game with occlusions: the participants and the observer. The participants are visible to and compete with each other, while the observer, such as a robot or an autonomous vehicle, observes the trajectories of visible (unoccluded) participants only. The observer's goal is to estimate both \textit{visible} and \textit{occluded} agents' objectives and trajectories based solely on observations of \textit{visible} agents. To this end, the observer seeks to estimate all agents' weighting parameters ${\textbf{w}}$ in the running cost \eqref{eqn:running cost} and trajectories ${\textbf{x}}$, which maximize the likelihood of the given noise-corrupted partial state observations (position only) from visible agents ${\textbf{y}^\mathcal{V}:=(\textbf{y}^i),i\in\mathcal{V}}$, i.e.,
\begin{subequations}
\begin{align}
\max_{\textbf{w},\textbf{x},\textbf{u}}\ &p(\textbf{y}^\mathcal{V}|\textbf{x},\textbf{u}),\label{eqn:likelihood}\\
\text{s.t.}\ &(\textbf{x},\textbf{u})\text{ OLNE of }\Gamma(\textbf{w},\textbf{x}_0,f),\label{eqn:nasheqconstraint}\end{align}
\label{eqn:MLE model}%
\end{subequations}
where $\mathcal{V}$ denotes the set of visible agents in the game.

We encode the Nash equilibrium condition \eqref{eqn:nasheqconstraint} by concatenating the first-order necessary (KKT) conditions: 
\begin{equation}
\begin{aligned}
\textbf{G}&(\textbf{x},\textbf{u},\pmb{\lambda}^i;\textbf{w}^i)\\&=\begin{bmatrix}
    \nabla_{\textbf{x}^i}\mathcal{L}^i(\textbf{x},\textbf{u},\pmb{\lambda}^i;\textbf{w}^i)\\
    \nabla_{\textbf{u}^i}\mathcal{L}^i(\textbf{x},\textbf{u},\pmb{\lambda}^i;\textbf{w}^i)\\
    x_{k+1}^i-f(x_k^i,u_k^i),\ k\in[T]
    \end{bmatrix}=\textbf{0},\ i\in[M],
\end{aligned}
\label{eqn:kkt}
\end{equation}
where $$\mathcal{L}^i(\textbf{x},\textbf{u},\pmb{\lambda}^i;\textbf{w}^i)=J^i(\textbf{x},\textbf{u};\textbf{w}^i)+\sum_{k=0}^{T-1}{\lambda_k^i}^\top\big(x_{k+1}^i-f(x_k^i,u_k^i)\big)$$ denotes each agent's Lagrangian with multipliers $\pmb{\lambda}^i$. Assuming that the observations are corrupted by white Gaussian noise:
$ {\textbf{y}_t^{\mathcal{V}}:=\begin{bmatrix}
    I_2 & \textbf{0}
\end{bmatrix}\textbf{x}_k^{\mathcal{V}}+\textbf{n}_k}, {\textbf{n}_k\sim\mathcal{N}(\textbf{0},\sigma^2I_2)}$, the maximum likelihood objective in (5a) becomes a least-square objective as follows:
\begin{subequations}
\begin{align}
\min_{\textbf{w},\textbf{x},\textbf{u},\pmb{\lambda}}\ &\frac{1}{|\mathcal{V}|\cdot T}\sum_{k=0}^{T-1}\sum_{j\in\mathcal{V}}\left\|y_k^j-\begin{bmatrix}
        I_2 & \textbf{0}\end{bmatrix}x_k^j\right\|_2^2, \label{eqn:squareerror}\\
\text{s.t.}\ &\textbf{G}(\textbf{x}^i,\textbf{u}^i,\pmb{\lambda}^i;\textbf{w}^i)=\textbf{0},\ i\in[M].\label{eqn:lse_kkt}
\end{align}
\label{eqn:LSE_model}%
\end{subequations}
The estimator then solves \eqref{eqn:LSE_model} to estimate all agents' weighting parameters $\textbf{w}$, trajectories $\textbf{x}$, and control sequences $\textbf{u}$ from the noise-corrupted observation of only visible agents' trajectories $\textbf{y}^{\mathcal{V}}$.

\subsection{Occlusion-Aware Contingency Game}
\label{sec:occlusion-aware contingency game}
Next, we outline the proposed occlusion-aware planning algorithm. A robot may be uncertain about the presence of potentially occluded agents during navigation. To manage this uncertainty, we construct a contingency game model akin to that of \cite{peters2024contingency}, which encodes both possibilities
with parameters ${\theta\in\Theta=\{\theta_1, \theta_2\}}$ and their corresponding belief likelihood $b(\theta)$, satisfying ${0\leq b(\theta_1)=1-b(\theta_2)\leq1}$.
We denote
\begin{equation}
    \begin{aligned}
    &\theta_1\text{: Occluded agents DO exist},\\
    &\theta_2\text{: Occluded agents DO NOT exist}.
    \end{aligned}
    \label{eqn:hypothesis}%
\end{equation}
A $M$-agent occlusion-aware contingency game is divided into two phases by the branching time $t_b$, after which it is assumed that any occluded agents will be visible.
For each time step $k < t_b$, the $i^\text{th}$ agent must choose a single control input which balances both hypotheses $(\theta_1, \theta_2)$. 
Because the $i^\text{th}$ agent knows that any occluded agents will be visible after $k = t_b$, planned control inputs for times $k \ge t_b$ are conditioned upon the value of $\theta$. The contingency game is thus denoted as $\Gamma_\text{con}(\Theta, t_b, \textbf{w}, \textbf{x}_0, f)$.

To encode this contingency structure, the $i^\text{th}$ agent constructs a larger game with two copies of each other agent: one for each hypothesis $\theta$. In this larger problem, the $j^\text{th}$ agent ($j\neq i$) is assumed to optimize
\begin{subequations}
    \begin{align}
    \min_{\textbf{x}^j_{\theta,i},\textbf{u}^j_{\theta,i}}\quad&J^j_{\theta,i}(\textbf{x},\textbf{u};\textbf{w}^j)
    \label{eqn:contingency_sub_costfunc}\\
    \text{s.t.}\quad &x_{k+1;\theta,i}^j=f(x_{k;\theta,i}^j,u_{k;\theta,i}^j), \quad k\in[T],\label{eqn:contingency_sub_dynamics}
    \end{align}
    \label{eqn:contingency_sub}%
\end{subequations}
where the $(\theta, i)$ subscripts indicate that this is the $j^\text{th}$ agent associated to observability condition $\theta$ in the $i^\text{th}$ agent's contingency game.
The $i^\text{th}$ agent then wishes to minimize its expected cost under distribution $b(\theta)$:
\begin{subequations}
    \begin{align}
    \min_{\textbf{x}^i_{\theta,i},\textbf{u}^i_{\theta,i}}\quad&\sum_{\theta\in\Theta}b(\theta)J_{\theta,i}^i(\textbf{x},\textbf{u};\textbf{w}^i)\label{eqn:contingency_main_costfunc}\\
    \text{s.t.}\quad &x_{k+1;\theta,i}^i=f(x_{k;\theta,i}^i,u_{k;\theta,i}^i), \ k\in[T],\label{eqn:contingency_main_dynamics}\\
    & u_{k;\theta_1,i}^i=u_{k;\theta_2,i}^i,\ k<t_b.
    \label{eqn:contingnecy_constraint}
    \end{align}
    \label{eqn:contingency_main}%
\end{subequations}
subject to the contingency constraint \eqref{eqn:contingnecy_constraint} that its control inputs in both interactions are identical for both $\theta_1$ and $\theta_2$ when $k<t_b$. When $k\geq t_b$ and the $i^{\text{th}}$ agent is certain about the (non)existence of occluded agents ($b(\theta_1)=1$ or $b(\theta_2)=1$),  it will pick the corresponding control input $u_{k;\theta,i}^i$. 

\begin{figure*}[!t]
\centering
\includegraphics[width=0.99\textwidth]{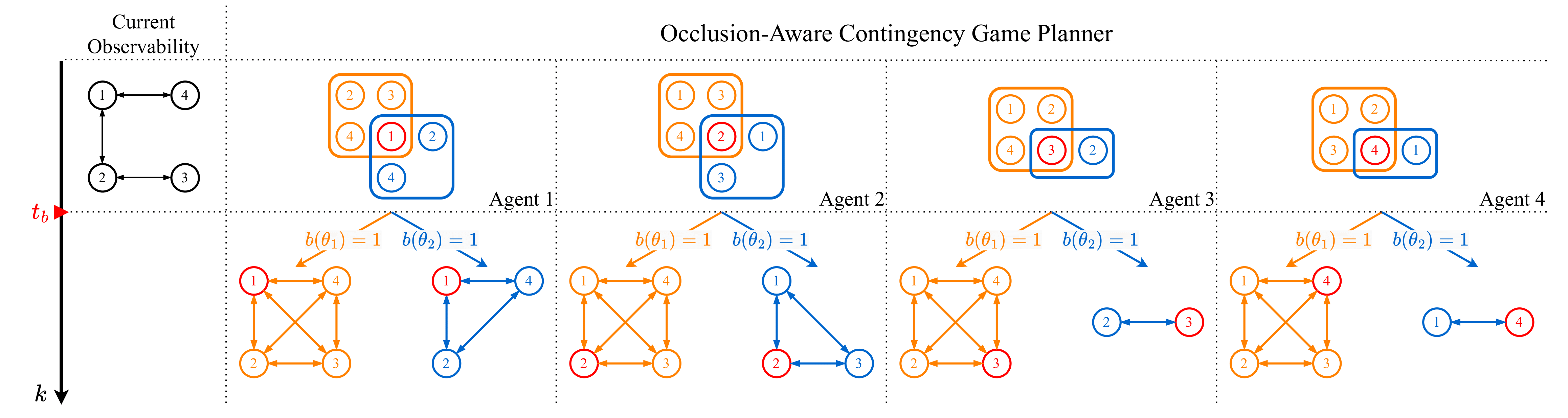}
\caption{4-agent occlusion-aware contingency game. When $k<\ ${\color{red}$t_b$}, agent $i$ considers two possibilities: {\color{orange}$\theta_1$ (occluded agents are present)} and {\color{NavyBlue}$\theta_2$ (there are no occluded agents)}. It chooses a single control input $u_k^i=\ ${\color{orange}$u_{k;\theta_1,i}^i$}$=\ ${\color{NavyBlue}$u_{k;\theta_2,i}^i$} balancing its uncertainty between both hypotheses $(${\color{orange}$\theta_1$}, {\color{NavyBlue}$\theta_2$}$)$.
When $k\geq\ ${\color{red}$t_b$}, any occluded agents are visible, and the $i^\text{th}$ agent will pick either {\color{orange}$u_{k;\theta_1,i}^i$} or {\color{NavyBlue}$u_{k;\theta_2,i}^i$} based on the ground truth value of $\theta$, i.e. the (non)existence of the occluded agents, which is assumed to be revealed at {\color{red}$t_b$}.}
\label{fig:contingency_planner}
\vspace{-3ex}
\end{figure*}

We use the following example to concretize the concept:\\
\textbf{Running Example}: Consider the perspective of Agent 1 (A1) in a 4-agent scenario illustrated in Figure \ref{fig:contingency_planner}. It plans its strategy in an occlusion-aware contingency game as follows:
\begin{itemize}
    \item $\theta_1$: A1 assumes that there is an occluded A3, resulting a 4-agent interaction (A1, A2, A3, A4).
    \item $\theta_2$: A1 assumes that there are only visible agents A2 and A4, resulting a 3-agent interaction (A1, A2, A4).
\end{itemize}
When $k<t_b$, A1 chooses a single control input $u_k^1$ in \eqref{eqn:contingnecy_constraint} balancing both hypotheses $(\theta_1, \theta_2)$. When $k\geq t_b$, A1 assumes it will know whether the occluded A3 exists or not, and can therefore choose its strategy based on the ground truth value of $\theta$.

\subsection{Receding Horizon Estimation and Planning}
\begin{algorithm}[!b]
\caption{Receding Horizon Estimation and Planning Pipeline in an Occlusion-Aware Contingency Game}
\begin{algorithmic}[1]
\renewcommand{\algorithmicrequire}{\textbf{Input:}}
\renewcommand{\algorithmicensure}{\textbf{Output:}}
\REQUIRE receding horizon game $\Gamma_{\text{RH}}(\textbf{w},\textbf{x}_0,f)$, receding horizon contingency game $\Gamma_\text{con,RH}(\Theta, t_b, \textbf{w}, \textbf{x}_0, f)$, trajectory observation $\textbf{y}^{\mathcal{V}_i}$, game horizon $T$, observation interval $K$, branching time $t_b$.
\ENSURE  Agent $i$'s control sequence in receding horizon\\ ${\textbf{u}_\text{RH}^i=\{u_{k|k}^{i}\}_{k=K}^\infty}$.
\FOR {$k=K$ to $\infty$}
\STATE {\color{blue}$\hat{\textbf{w}}^{-i},\hat{\textbf{x}}_k^{-i}$}$\leftarrow$ solving the maximum likelihood problem given in \eqref{eqn:recedinghorizonmlemodel}. {\color{blue}[Estimation]}
\STATE {\color{red}$u_{k|k}^{i*}$}$\leftarrow$ solving $\Gamma_\text{con,k}(\Theta, t_b, \textbf{w}^i,${\color{blue}$\hat{\textbf{w}}^{-i}$}$, x_k^i,${\color{blue}$\hat{\textbf{x}}_k^{-i}$}$,f)$ given by \eqref{eqn:contingency_sub} and \eqref{eqn:contingency_main}. {\color{red}[Planning]}
\STATE $x_{k+1}^i\leftarrow f(x_k^i,${\color{red} $u_{k|k}^{i*}$}$)$ by state update \eqref{eqn:nashdynamics}.
\ENDFOR
\end{algorithmic}
\label{alg:algorithm}
\end{algorithm}

We now integrate the occlusion-aware game estimator in Section \ref{sec:occluded_agent_behavior_inference} and the occlusion-aware contingency game planner in Section \ref{sec:occlusion-aware contingency game}. This enables the robot to estimate the states of both visible and occluded agents from noisy position observations and then plan its own control input in a receding horizon fashion based on the estimated states.\\ 

\noindent\textbf{Receding Horizon Game}: We denote an $M$-agent receding horizon game by $\Gamma_{\text{RH}}(\textbf{w},\textbf{x}_0,f):=\{\Gamma_k(\textbf{w},\textbf{x}_k,f)\}_{k=0}^\infty$, consisting of games with initial states $\textbf{x}_k$ over all time  steps $k$. At each step, the $i^{\text{th}}$ agent solves for an OLNE strategy $(u_{k|k}^{i},\dots,u_{k+T-1|k}^{i})$ for ${\Gamma_k(\textbf{w},\textbf{x}_k,f)}$, It then implements the first control input $u_{k|k}^{i}$ and updates its state according to \eqref{eqn:nashdynamics}. The resulting control sequence is denoted ${\textbf{u}_{\text{RH}}^{i}=\{u_{0|0}^{i*},u_{1|1}^{i*},\cdots\}}$. 
We similarly define the receding horizon contingency game as ${\Gamma_\text{con,RH}(\Theta, t_b, \textbf{w}, \textbf{x}_0, f):=\{\Gamma_\text{con,k}(\Theta, t_b, \textbf{w}, \textbf{x}_0, f)\}_{k=0}^\infty}$.\\

\noindent\textbf{Receding Horizon Open-Loop Nash Equilibrium}: We define ${\textbf{u}_{\text{RH}}^{i*}:=\{u_{0|0}^{i*},u_{1|1}^{i*},\cdots\}}$ as the receding horizon Nash strategy for $\Gamma_{\text{RH}}(\textbf{w}, \textbf{x}_0, f)$ if $\forall k, u_{k|k}^{i*}$ is an OLNE strategy to $\Gamma_k(\textbf{w},\textbf{x}_k,f)$, with ${\textbf{x}_{\text{RH}}^{i*}:=\{x_{1|0}^{i*},x_{2|1}^{i*},\cdots\}}$ being the corresponding Receding Horizon Open-Loop Nash Equilibrium (RHOLNE) trajectory of the $i^{\text{th}}$ agent.\\

\noindent\textbf{Receding Horizon Estimation and Planning}: We summarize the receding horizon estimation and planning pipeline in Algorithm \ref{alg:algorithm}. At each time $k$, each agent estimates the parameters and states of the other agents from the previous $K$ observations $\{\textbf{y}_t^{\mathcal{V}_i}\}_{t=k-K}^k$, i.e.,
\begin{subequations}
\begin{align}
\max_{\textbf{w},\textbf{x}_{\text{RH}},\textbf{u}_{\text{RH}}}\ &p(\{\textbf{y}_t^{\mathcal{V}_i}\}_{t=k-K}^k|\textbf{x}_{\text{RH}},\textbf{u}_{\text{RH}}),\label{eqn:recedinghorizonmle}\\
\text{s.t.}\ &(\textbf{x}_{\text{RH}},\textbf{u}_{\text{RH}}) \text{ RHOLNE of }\Gamma_{\text{RH}}(\textbf{w},\textbf{x}_0,f),\label{eqn:recedinghorizonmleconstraint}\end{align}
\label{eqn:recedinghorizonmlemodel}%
\end{subequations}
where $\mathcal{V}_i$ denotes the set of agents that are visible to the $i^{\text{th}}$ agent. The agent then plans its strategy by solving the contingency game ${\Gamma_\text{con,k}(\Theta, t_b, \textbf{w}, \textbf{x}_0, f)}$ based on the estimated states of the other agents and implements the first control input $u_{k|k}^{i*}$.

In the next section, we conduct simulation experiments to evaluate the estimation performance of our method under observation noise and the receding horizon planning pipeline's contribution to safer decision-making in driving scenarios.

%% file: experiment.tex
\label{sec:experiments}
To evaluate our method's performance in promoting safety in autonomous driving scenarios under occlusions, in this section, we conduct 1) a 3-agent interaction experiment to evaluate our methods' estimation performance, 2) a multi-agent planning experiment to evaluate our methods' collision avoidance performance, and 3) a crossing-road experiment to evaluate our method's performance in estimation and planning in driving scenarios.
\subsection{3-Agent Interaction (Estimation)}
\label{sec:three_player_interaction}
\begin{figure*}[!t]
    \centering
    \subfloat[Ground truth]{\includegraphics[width=0.243\textwidth]{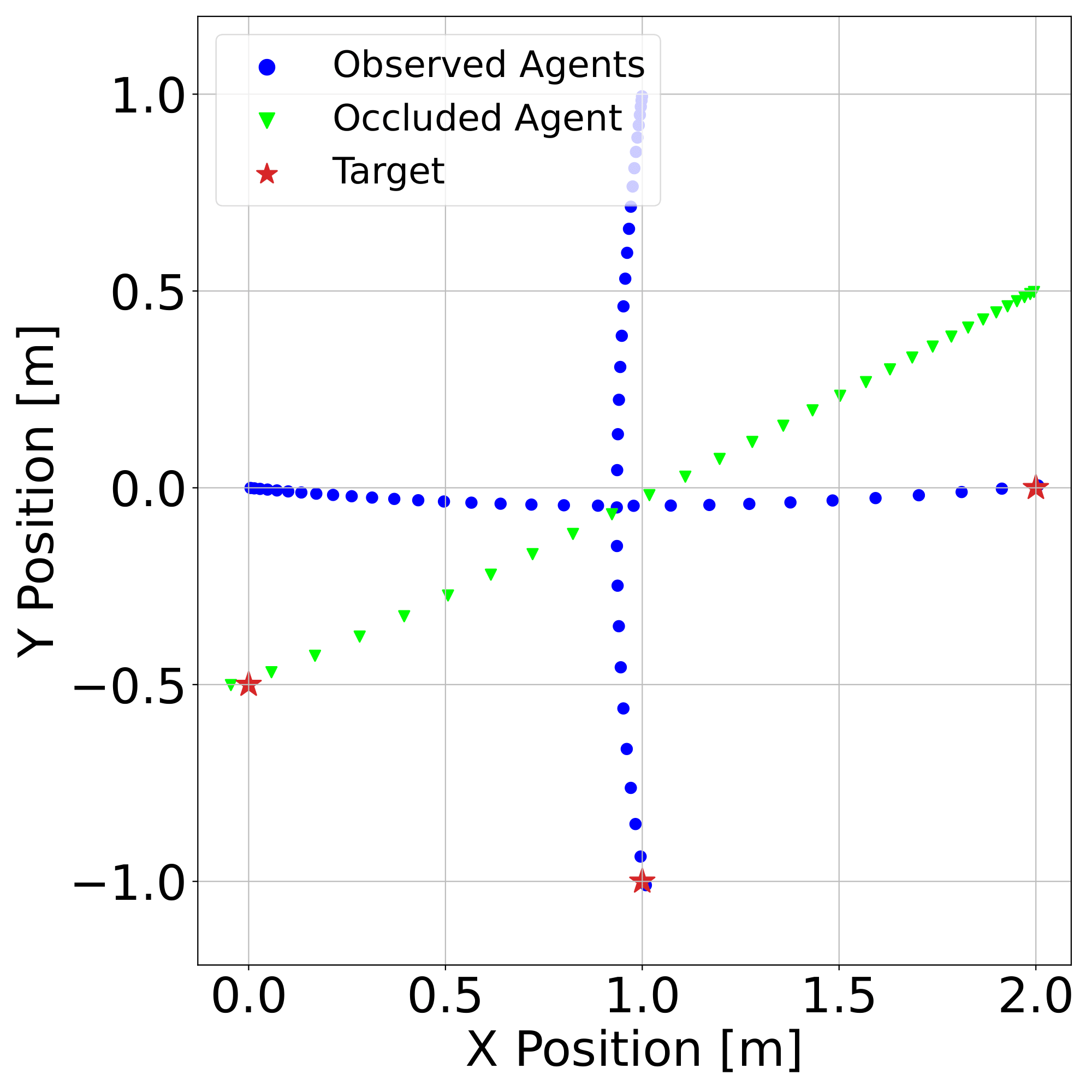}
    \label{fig:demo_gt}}
    \subfloat[Noisy Observations]{\includegraphics[width=0.243\textwidth]{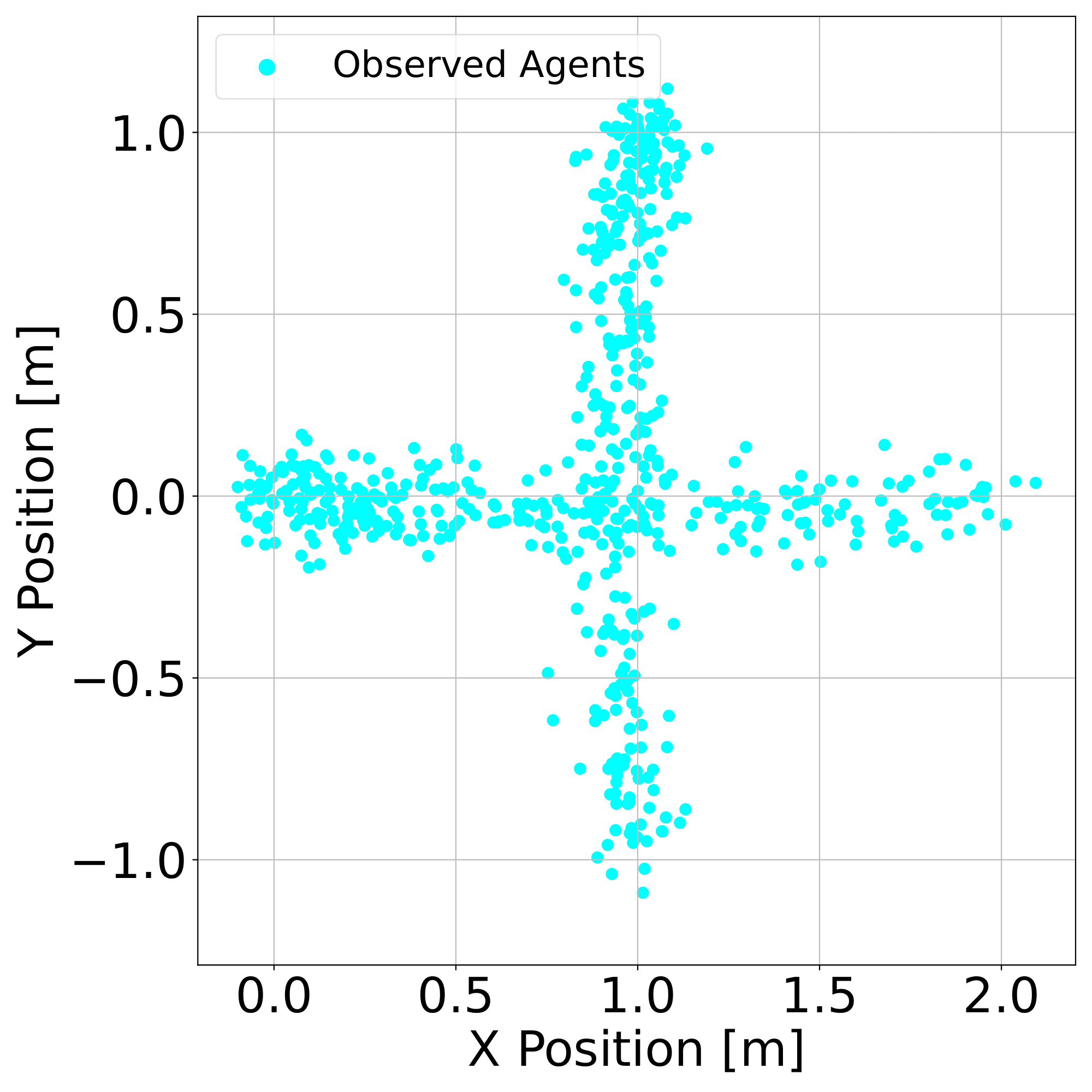}
    \label{fig:demo_obs}}
    \subfloat[Estimation (Occlusion-aware)]{\includegraphics[width=0.243\textwidth]{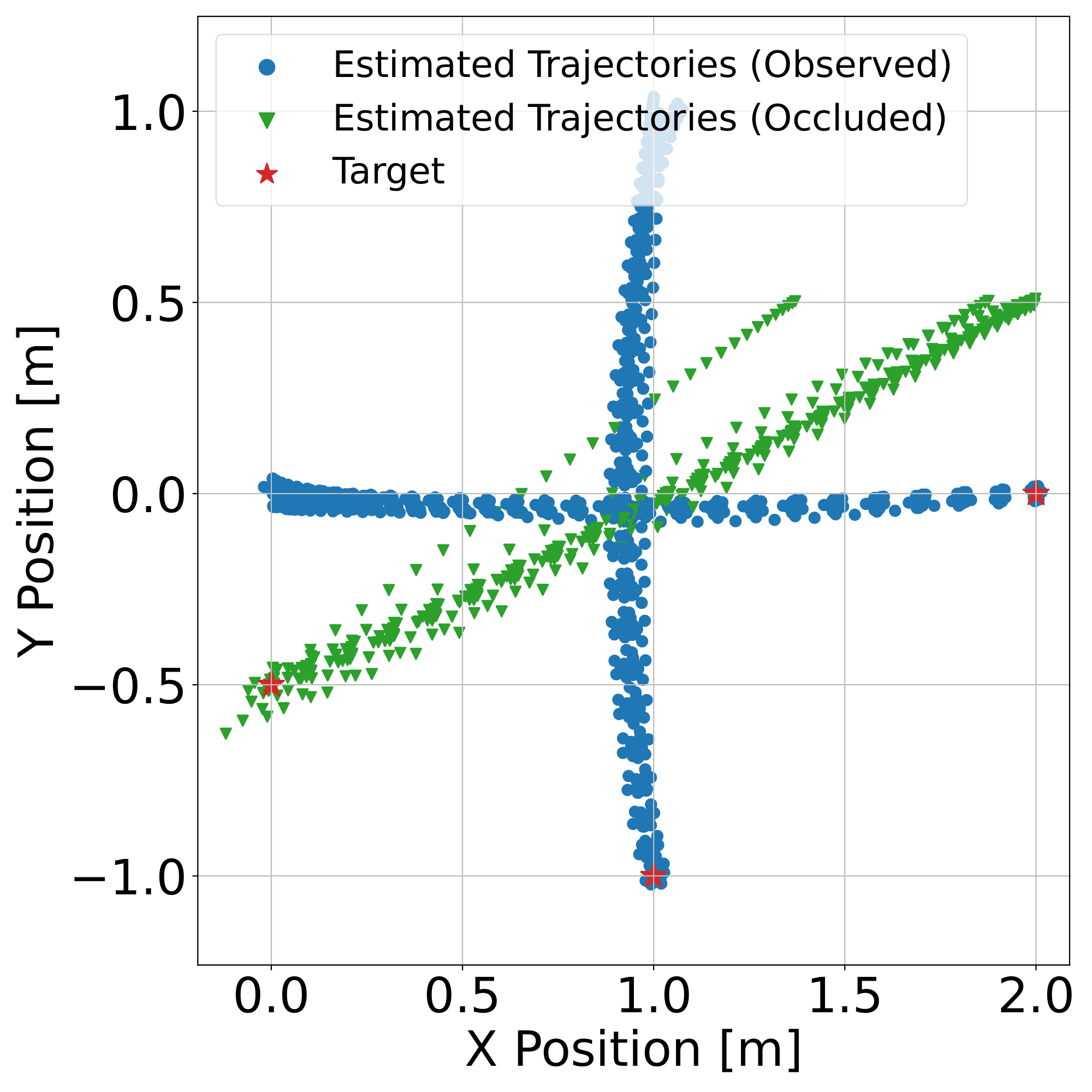}
    \label{fig:demo_est_aware}}
    \subfloat[Estimation (Occlusion-ignorant)]{\includegraphics[width=0.243\textwidth]{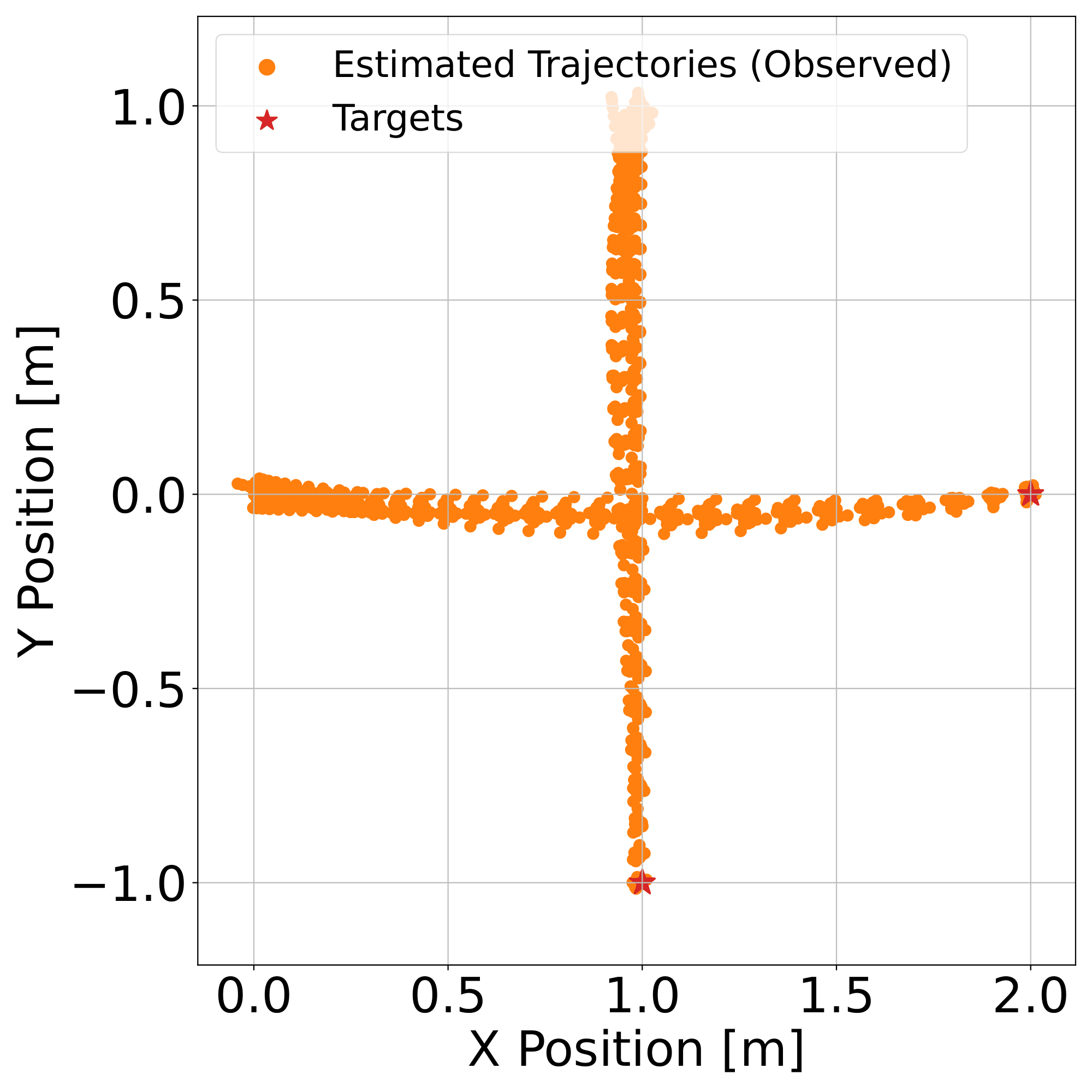}
    \label{fig:demo_est_ignorant}}
    \caption{Demonstration for the 3-agent interaction scenario in Section \ref{sec:three_player_interaction}, with observation noise standard devioveration $\sigma=0.07$\si{m} across 20 different observation sequences, in which all agents apply open-loop Nash strategies. (a)  Ground truth trajectory of each agent. (b)  Observations, where the trajectories of the blue agents are corrupted by noise and the green agent is occluded. (c)  Estimated trajectories by the occlusion-aware game estimator (\textbf{ours}). (d) Estimated trajectories by the occlusion-ignorant estimator (\textbf{baseline}). Our method 1) enables estimation of the occluded agent, and 2) provides more accurate estimation of observed agents than the baseline.}
    \label{fig:demo_and_estimation}
    \vspace{-3.5ex}
\end{figure*}

\begin{figure}[!t]
\centering
\subfloat[Parameter Estimation Performance]{\includegraphics[width=0.479\textwidth]{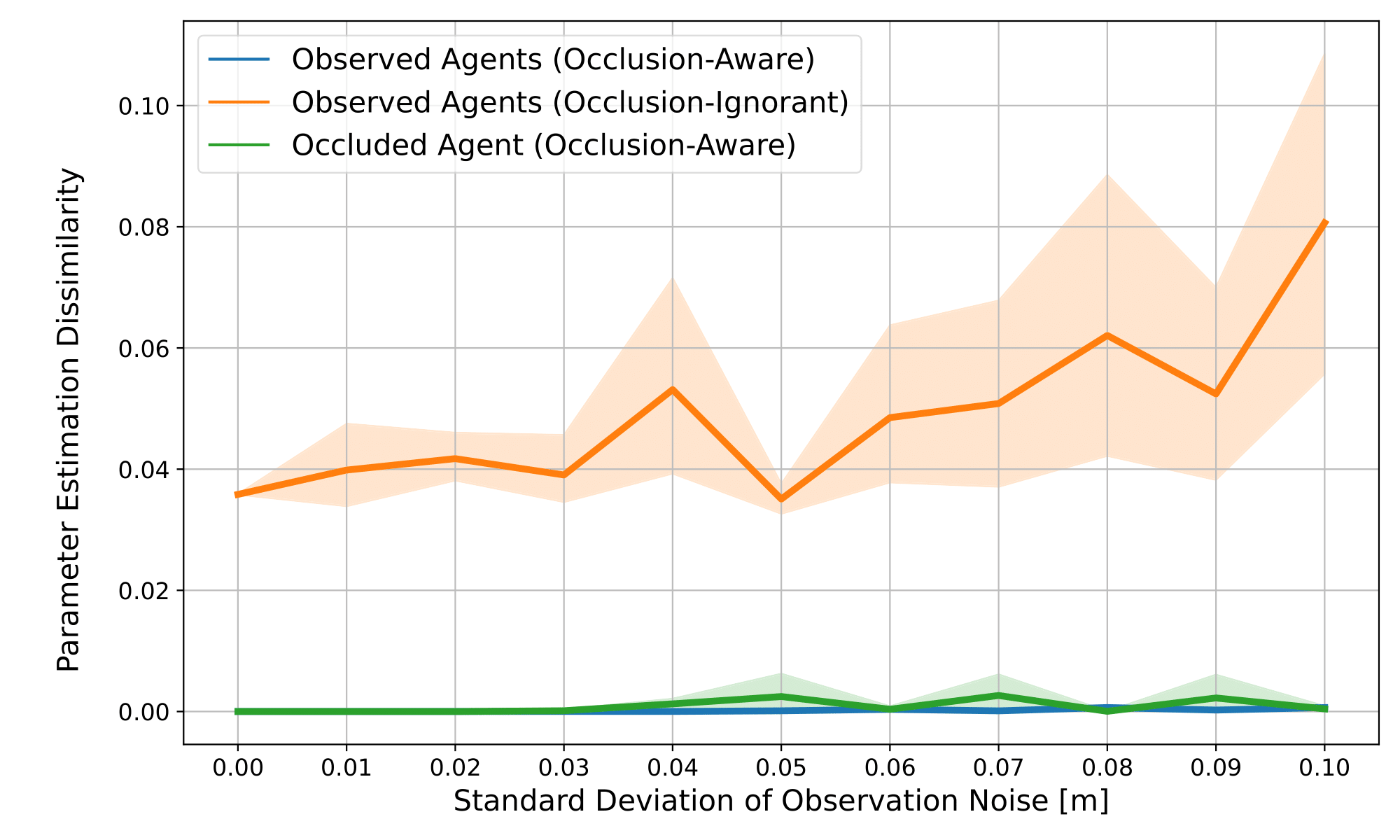}
\label{fig:parameter_estimation_performance}}\vspace{-2.5ex}

\subfloat[Trajectory Estimation Performance]{\includegraphics[width=0.479\textwidth]{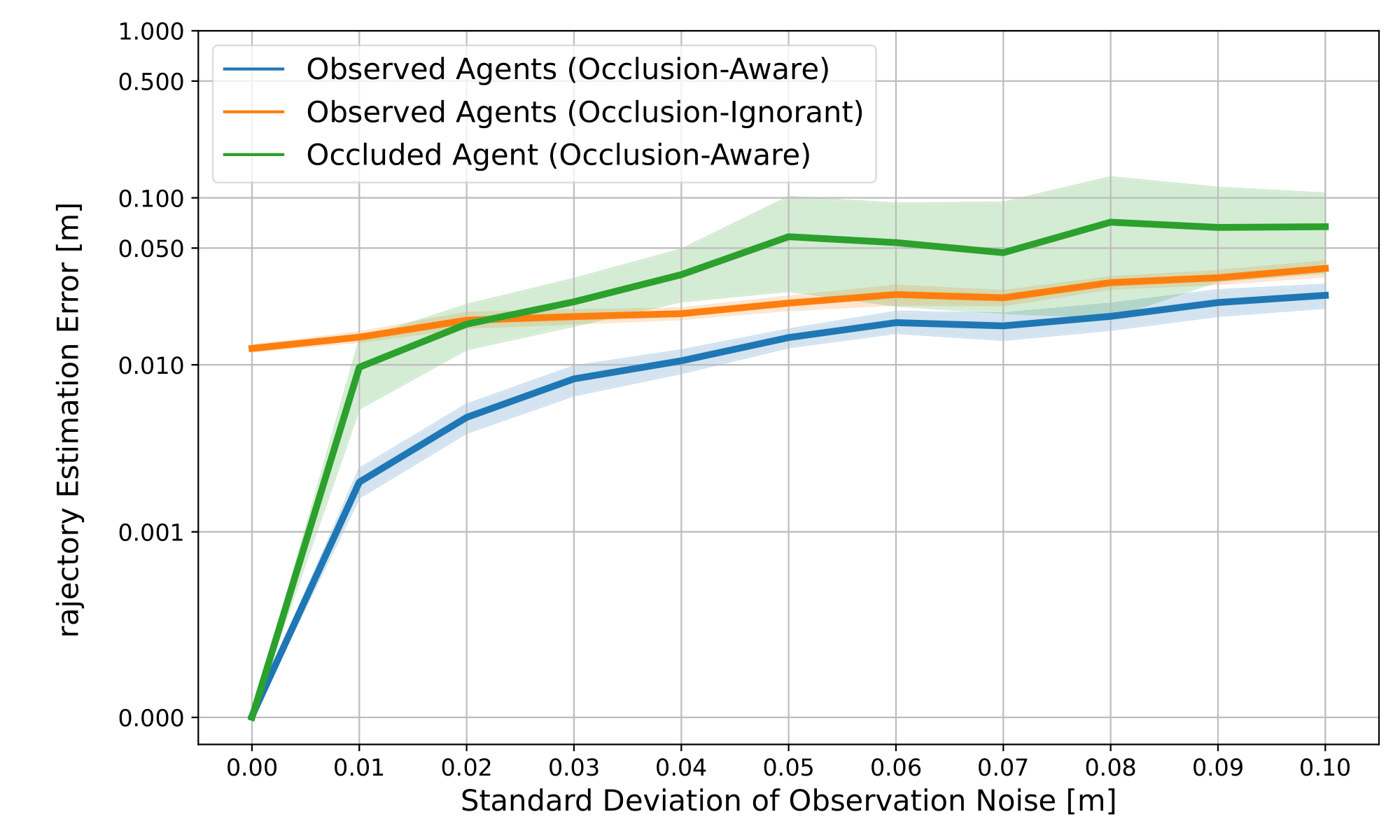}
\label{fig:trajectory_estimation_performance}}
\caption{Estimation performance for occlusion-aware game estimator (\textbf{ours}) and the occlusion-ignorant game estimator (\textbf{baseline}). (a) Parameter estimation performance. (b) Trajectory estimation performance. Our method estimates the weighting parameters and state trajectories of both visible and occluded agents more accurately than the baseline.}
\label{fig:estimation_analysis}
\vspace{-4ex}
\end{figure}

We first evaluate the estimation performance in a 3-agent interaction scenario, where the robot observes an interaction with two visible agents and one occluded agent with the cost structure in \eqref{eqn:running cost}. While all three agents are visible to each other, the robot has noisy partial observations of only the visible agents, as depicted in Figures \ref{fig:demo_gt} and \ref{fig:demo_obs}.\\
\noindent\textbf{Baseline:} We compare our method with an occlusion-ignorant method, i.e., the robot is ignorant of the occluded agent and identifies the interaction as a 2-agent game.\\
\noindent\textbf{Evaluation Metrics}: We evaluate the estimation performance with the following metrics:\\
\textit{Weighting Parameter Estimation Precision}: We evaluate the parameter cosine dissimilarity of each agent \cite{peters2021inferring}:
\begin{equation}
    D(\textbf{w}^i_{\text{GT}},\hat{\textbf{w}}^i):=1-\frac{{\textbf{w}_{\text{GT}}^i}^\top \hat{\textbf{w}}^i}{\|\textbf{w}_{\text{GT}}^i\|_2\|\hat{\textbf{w}}^i\|_2},\ i\in[M]
    \label{eqn:parametererror}
\end{equation}
which quantifies the precision of estimated weighting parameters $\hat{\textbf{w}}$ in comparison to the ground truth $\textbf{w}_{\text{GT}}$.\\
\textit{Trajectory Estimation Accuracy}: We compute the average displacement error (ADE) between the estimated trajectories based on recovered parameters and the ground truth:
\begin{equation}
    \begin{aligned}
    \text{ADE}_{\text{visible}}&:=\frac{1}{|\mathcal{V}|\cdot T}\sum_{i\in\mathcal{V}}\sum_{k\in[T+1]}\|p_{\text{GT},k}^i-\hat{p}_{k}^i\|_2,\\
    \text{ADE}_{\text{occluded}}&:=\frac{1}{|\mathcal{O}|\cdot T}\sum_{i\in\mathcal{O}}\sum_{k\in[T+1]}\|p_{\text{GT},k}^i-\hat{p}_{k}^i\|_2,
    \end{aligned}
    \label{eqn:positionerror}
\end{equation}
where ${p_{\text{GT},k}^i}$ and ${\hat{p}_{k}^i}$ denote the ground truth and the estimated position of the $i^{\text{th}}$ agent at time step $k$, respectively.\\
\textbf{Discussion}: Monte Carlo study results of 20 observation sequences across 11 levels of observation noise reveal that our method 1) outperforms the baseline that ignores the occluded agent in estimating the parameters and trajectories of the observed agents, and 2) is able to estimate those of the occluded agent as well, as illustrated in Figure \ref{fig:estimation_analysis}. Figure \ref{fig:demo_and_estimation} visualizes the estimated trajectories for our approach and the baseline; as shown, our method reliably recovers the occluded agent’s trajectory while also maintaining high accuracy for visible agents.

\subsection{Collision Avoidance Scenario (Planning)}
\begin{figure}[!t]
\centering
\subfloat[Contingency Game Planner]{\includegraphics[width=0.236\textwidth]{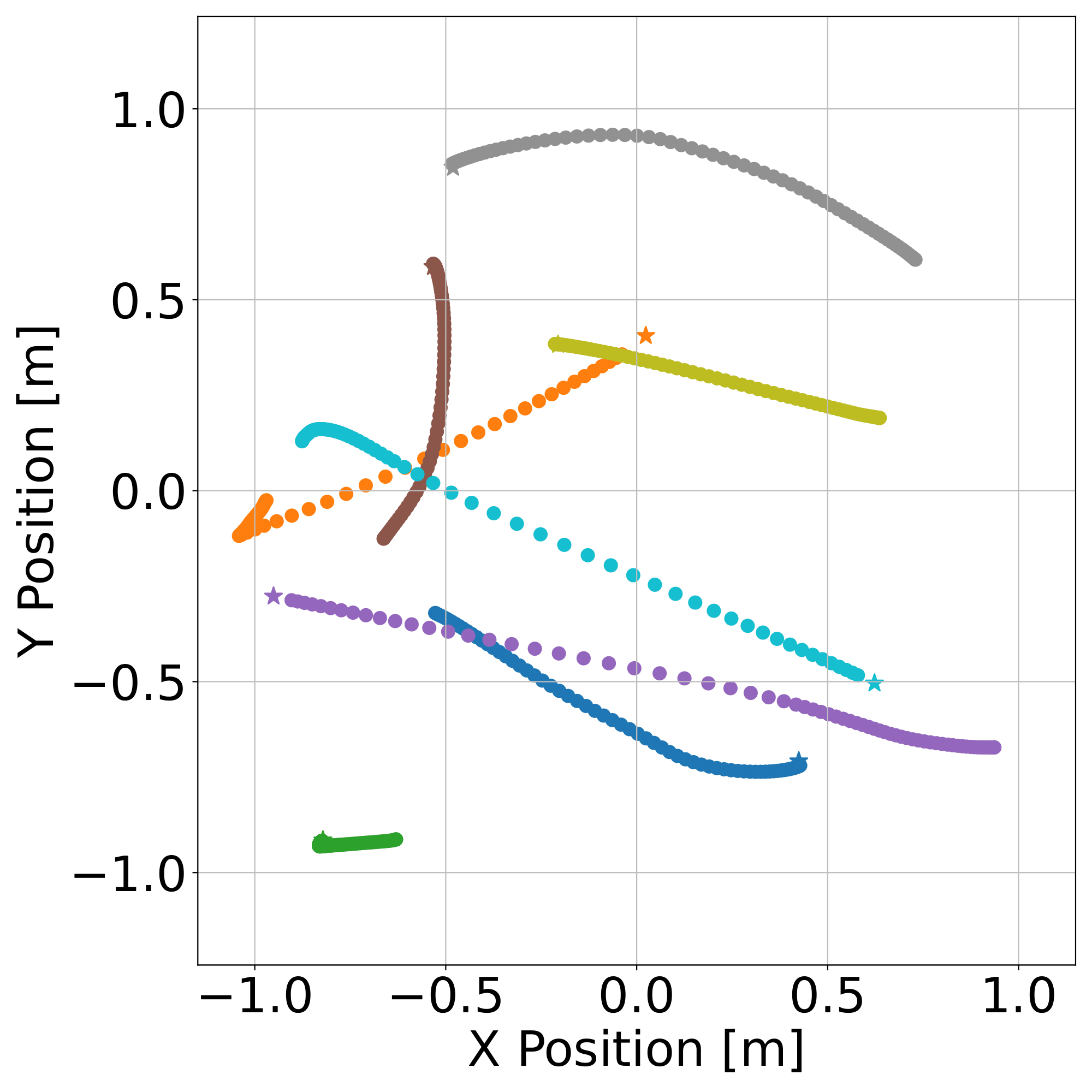}
\label{fig:contingnecy_planning}}
\subfloat[Occlusion-Ignorant Game Planner]{\includegraphics[width=0.236\textwidth]{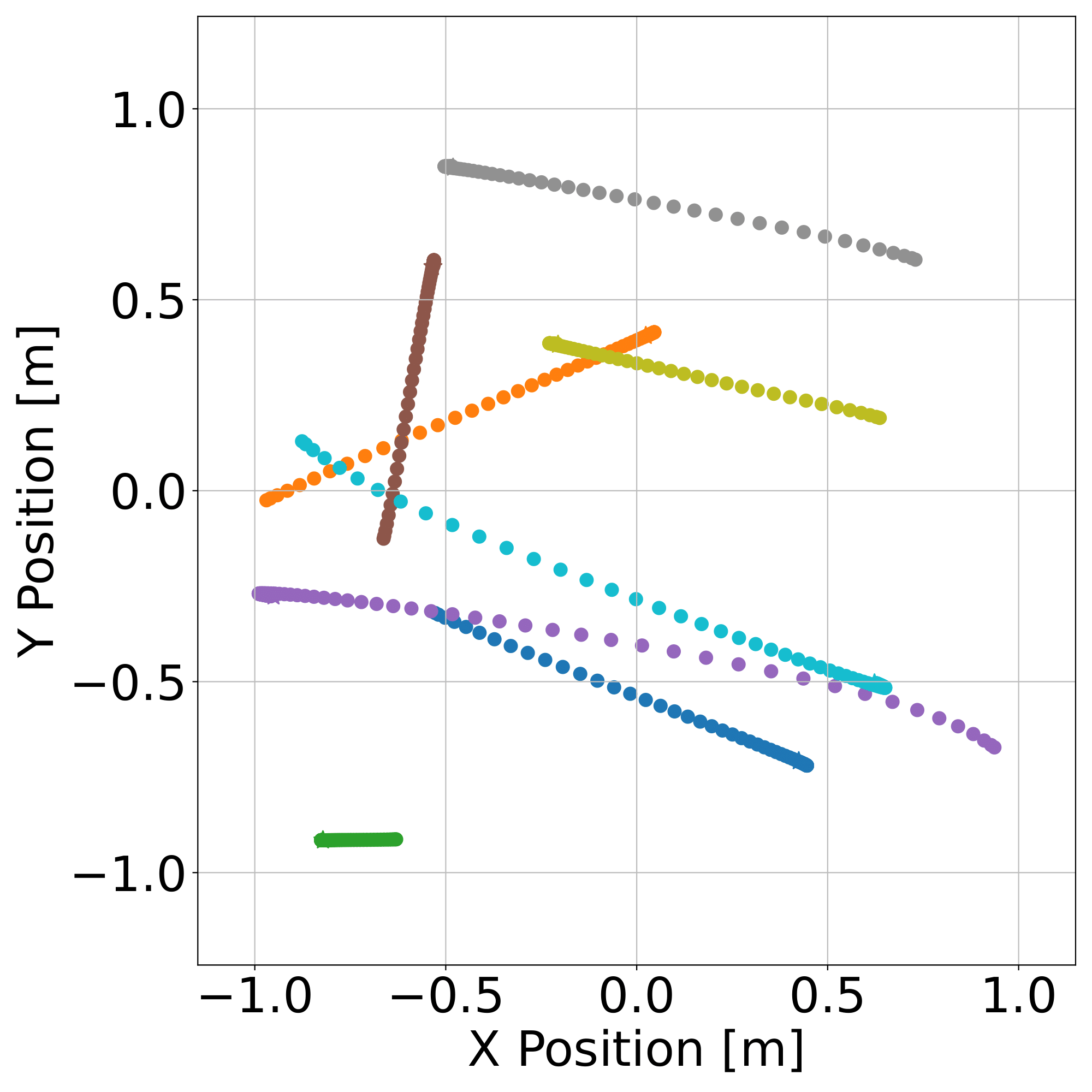}
\label{fig:occlsion_ignorant_planning}}
\caption{Trajectories of agents in an 8-agent planning scenario using (a) contingency game planner (\textbf{ours}) with belief $b(\theta_1)=0.7$ and branching time $t_b=4$\si{s}; and (b) occlusion-ignorant game planner (\textbf{baseline}) with branching time $t_b=2$\si{s}. Our method enables agents to reason proactively in order to avoid potential collisions with occluded agents.}
\label{fig:planning_visualization}
\vspace{-4ex}
\end{figure}

Next, we evaluate the performance in a collision avoidance scenario, where each agent aims to reach its target while keeping as far as possible from the other agents (using the running cost in \eqref{eqn:running cost}). Specifically, the visibility among agents is randomly generated, and each agent cannot see the occluded agents until the branching time $t_b$. In order to isolate the effects of planning, we do not include observation noise in this experiment.\\
\textbf{Baseline:} We compare our method with an occlusion-ignorant planner, i.e., before $t_b$, each agent plans based only on the agents visible to him, and after $t_b$, they plan based on all agents.\\
\noindent\textbf{Evaluation Metric:} We evaluate the collision avoidance performance with the following metric:\\
\textit{Minimum Distance}: We evaluate the minimum distance between the $i^\text{th}$ agent and all other agents over time:\\
\begin{equation}
    d_{\text{min}}:=\min_{k\in[T+1]}\min_{\substack{i,j\in[M]\\j\neq i}}\ {\|p_k^i-p_k^j\|}_2.
\label{eqn:minimumdist}
\end{equation}
\noindent\textbf{Discussion:} We compare our method (with different values of belief $b(\theta_1)$ and branching time $t_b$) to the occlusion-ignorant baseline. Monte Carlo study results of 100 samples (goals and initial positions randomly generated) in the 4, 6, and 8-agent scenarios reveal that the agents behave more conservatively before $t_b$. Consequently, a larger $t_b$ enables agents to maintain larger distances from both visible and occluded agents. With the same $t_b$, variations in $b(\theta_1)$ do not have a significant effect. Across all values of $b(\theta_1)$ and $t_b$, our method outperforms the occlusion-ignorant baseline in collision avoidance. Specifically, agents equipped with the occlusion-aware contingency game planner make more careful decisions and avoid the other agents more actively, as illustrated in Figure \ref{fig:planning_visualization}.
\begin{figure}[!t]
\centering
\includegraphics[width=0.486\textwidth]{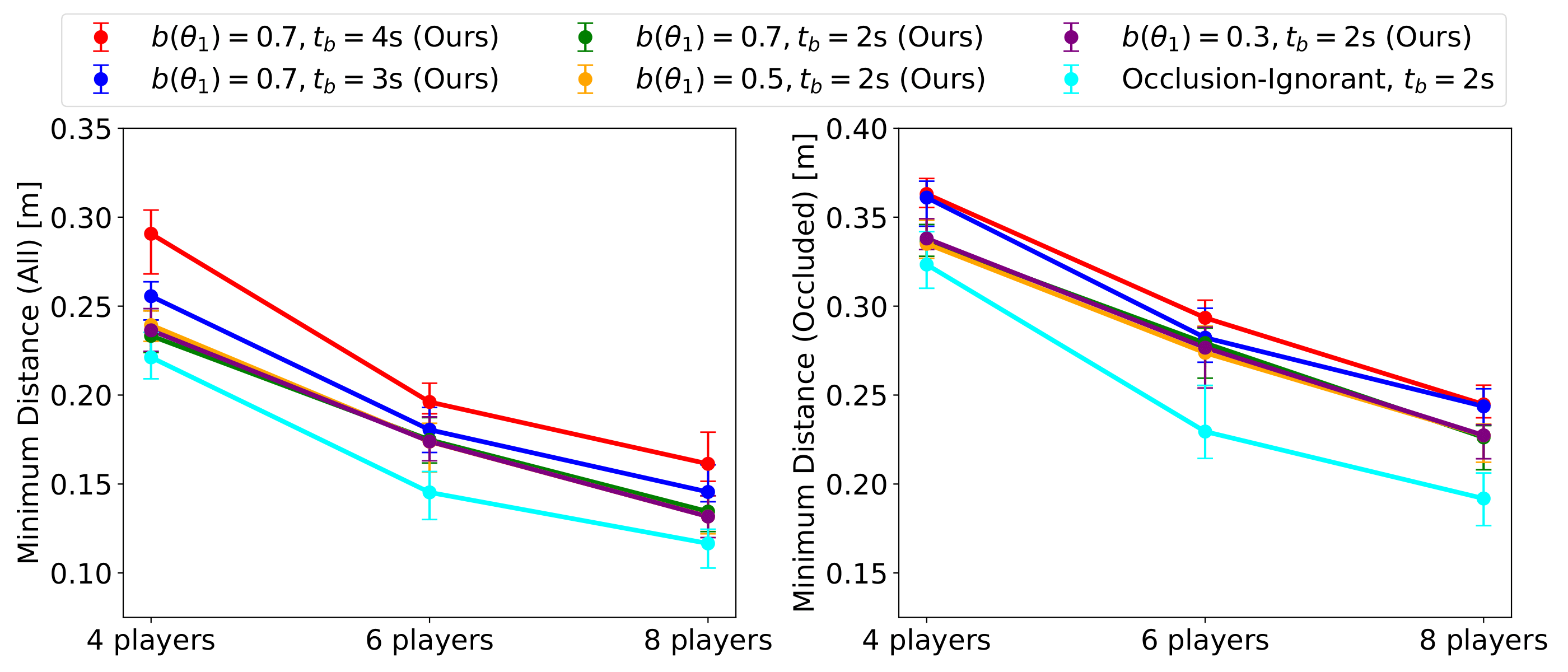}
\caption{Bootstrapped confidence intervals of the median for $d_\mathrm{min}$ \eqref{eqn:minimumdist} 
in 4, 6, and 8-agent planning scenarios. Left: Minimum distance between all pairs of agents. Right: Minimum distance between agents that are initially occluded to each other. Our method enables the agents to keep a larger distance from both visible and occluded agents than the occlusion-ignorant baseline, and this pattern persists across multiple values of $b(\theta_1)$ and $t_b$.}
\label{fig:planning_analysis}
\vspace{-4ex}
\end{figure}

\subsection{Crossing-Road Scenario (Estimation and Planning)}
\begin{figure*}[!t]
    \includegraphics[width=0.98\textwidth]{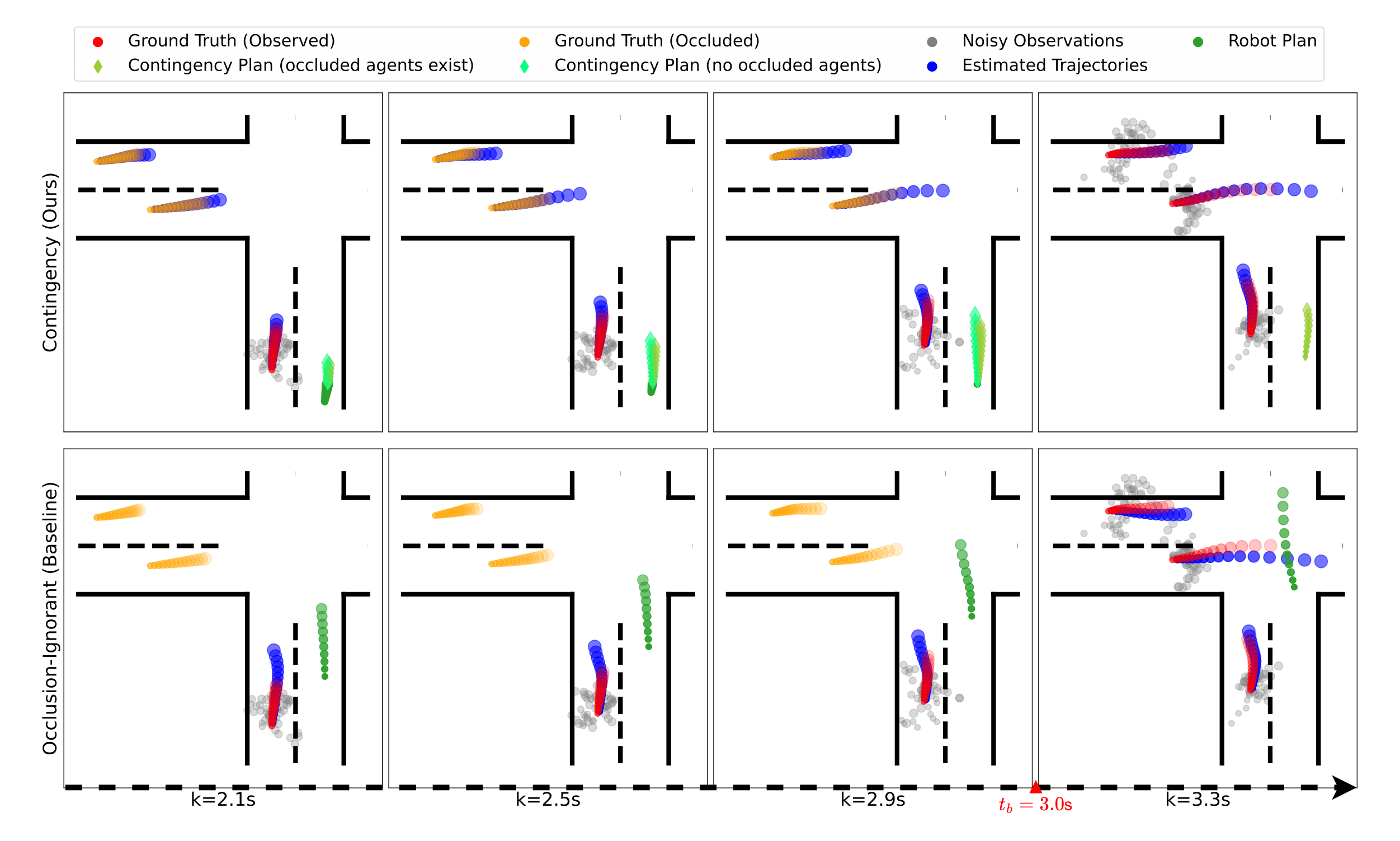}
    \caption{Crossing road scenario. Top: the green vehicle is equipped with the occlusion-aware game estimator and contingency game planner. When $k<t_b$, the green vehicle estimates all agents' trajectories from noisy observation of the red vehicle next to it and formulates a contingency plan which accounts for the (non)existence of the occluded vehicles on the horizontal lanes. When $k\geq t_b$, occlusions are removed and the green vehicle follows the appropriate branch of its contingency plan. 
    Bottom: the green vehicle is equipped with the occlusion-ignorant game estimator and planner. It ignores the occluded vehicles until $k\geq t_b$. Our method enables the green vehicle to recognize the risk posed by potentially occluded vehicles, and slow down rather than blindly entering the intersection and causing a near-accident.
    }
    \label{fig:traffic_plot}
    \vspace{-2ex}
\end{figure*}
In this scenario, we consider a realistic crossing-road scenario with observation noise. The running cost of each agent is formulated as follows:
\begin{equation}
    g_k^i(\textbf{x}_k,\textbf{u}_k;\textbf{w}^i)=\sum_{l=1}^4w_l^ig_{l,k}^i\left\{\begin{aligned}
    &g_{1,k}^i=\|p_k^i-p_{g}^i\|_2^2\\
    &g_{2,k}^i=\sum_{\substack{j=1\\j\neq i}}^M\frac{1}{\|p_k^i-p_k^{j}\|_2^2}\\
    &g_{3,k}^i=(d_k^i)^2\\
    &g_{4,k}^i=\|u_k^i\|_2^2
    \end{aligned}\right.,
\end{equation}
Here $d_k^i$ denotes the distance between the $i^{\text{th}}$ agent and the center line of the lane it belongs to, and the goal of each agent $x_{g,k}^i$ evolves with time. In this scenario, each agent not only tries to keep a safe distance from other agents ($g_{1,k}^i$) while moving forward ($g_{2,k}^i$) but also seeks to keep itself in the lane according to traffic rules ($g_{3,k}^i$) without heavy energy consumption ($g_{4,k}^i$).\\
\noindent\textbf{Baseline:} We compare our method with a receding horizon Nash game-theoretic planner equipped with the occlusion-ignorant estimator.\\
\noindent\textbf{Evaluation Metrics:} We evaluate the \textit{Trajectory Estimation Accuracy} \eqref{eqn:positionerror} for estimation performance, and the \textit{Minimum Mutual Distance} \eqref{eqn:minimumdist} for planning performance.\\
\noindent\textbf{Discussion:} Figure \ref{fig:traffic_plot} illustrates the difference between our method and the baseline. When $k<t_b$\footnote{We keep a constant $t_b$ for both methods for a fair comparison.}, the orange vehicles are occluded from the green vehicle. The baseline method assumes that the green vehicle only interacts with the red observed vehicle and plans in a 2-agent game. Our method, with the occlusion-aware game estimator, estimates all agents' states and plans in a 4-agent contingency game. The occlusion-aware game estimator infers the trajectories of occluded vehicles, which cause the (visible) red vehicle to decelerate; then the contingency game planner provides a more conservative strategy before the branching time $t_b$. When $k\geq t_b$, the occluded vehicles come into view. The baseline method now solves a 4-agent game, while our method chooses the strategy that assumes occluded vehicles exist. Statistical results in Table \ref{tab:trafficstats} indicate that our method provides 1) accurate estimation of occluded vehicles, 2) more accurate estimations of all vehicles, and 3) a safer collision avoidance strategy that keeps 1\si{m} farther away from initially occluded vehicles in comparison with the baseline.

\begin{table}[!t]
  \centering
  \scriptsize
  \begin{threeparttable}
  \caption{Performance in the crossing-road scenario}
    \begin{tabular}{ccc}
    \toprule
    \toprule
          & occlusion-aware estimator & occlusion-ignorant game \\ 
          & contingency game planner & estimator and planner \\
          & (\textbf{ours}) & (\textbf{baseline}) \\
    \midrule
    $\text{ADE}_\text{observed}$ [\si{m}] & \textbf{0.41} / \textbf{0.58} / 1.08 & 0.57 / 0.69 / \textbf{0.99} \\
    $\text{ADE}_\text{occluded}$ [\si{m}] & \textbf{0.51} / \textbf{0.80} / \textbf{1.18} &  - \tablefootnote{The baseline cannot estimate occluded agents' trajectories.}\\
    $d_{\text{min, observed}}$ [\si{m}] & \textbf{1.35} / \textbf{1.36} / \textbf{1.42} & 1.25 / 1.27 / 1.32 \\
   $d_{\text{min, occluded}}$ [\si{m}] & \textbf{4.61} / \textbf{4.67} / \textbf{4.68} & 3.66 / 3.70 / 4.58 \\
    \bottomrule
    \bottomrule
    \end{tabular}%
  \label{tab:trafficstats}%
  \end{threeparttable}
  \vspace{-5ex}
\end{table}%

%% file: conclusion.tex
In this work, we introduced a novel occlusion-aware behavior modeling approach to infer the behavior of occluded agents from noise-corrupted observations of only observed agents. Our method recovers parameters of an underlying game model that best explains the observed trajectories and simultaneously computes open-loop Nash trajectories for both visible and occluded agents. We integrated our inference method with a contingency game planner which operates in a receding horizon fashion to make robot decisions in navigation tasks. Experimental results underscore our approach's advantage in providing accurate estimates of the game model and agents' trajectories over the occlusion-ignorant baseline. Meanwhile, our planning algorithm also outperforms a state-of-the-art but occlusion-ignorant baseline, revealing its potential for proactively avoiding collisions with occluded agents during navigation.

While our evaluation was conducted in simulated environments, our future research will focus on extending our approach to incorporate real sensor data, such as lidar and camera observations, for deployment in realistic urban traffic scenarios and social navigation. This evolution promises to enhance the feasibility of our method in real-world autonomous navigation and robotic systems.